\documentclass[twocolumn,showpacs,preprintnumbers,amsmath,amssymb]{revtex4}

\usepackage{graphicx}
\usepackage{dcolumn}
\usepackage{bm}
\usepackage{latexsym}
\usepackage{amsfonts}
\usepackage{amssymb}
\usepackage{amsmath}

\begin{document}
\preprint{KUNS-1976}
\title{Braneworld Flux Inflation }

\author{Sugumi Kanno}
\email{sugumi@tap.scphys.kyoto-u.ac.jp}
\author{Jiro Soda}
\email{jiro@tap.scphys.kyoto-u.ac.jp}
\author{David Wands}
\email{david.wands@port.ac.uk}
\affiliation{Institute of Cosmology \& Gravitation, University of Portsmouth,
 Portsmouth, PO1 2EG, United Kingdom}
\affiliation{
 Department of Physics, Kyoto University, Kyoto 606-8501, Japan
}%

\date{\today}

\begin{abstract}
  We propose a geometrical model of brane inflation where inflation is
  driven by the flux generated by opposing brane charges and
  terminated by the collision of the branes, with charge annihilation.
  We assume the collision process is completely inelastic and the
  kinetic energy is transformed into the thermal energy after
  collision. Thereafter the two branes coalesce together and behave as a
  single brane universe with zero effective cosmological constant.  In
  the Einstein frame, the 4-dimensional effective theory changes
  abruptly at the collision point. Therefore, our inflationary model
  is necessarily 5-dimensional in nature.  As the collision process
  has no singularity in 5-dimensional gravity, we can follow the
  evolution of fluctuations during the whole history of the universe.
  It turns out that the radion field fluctuations have a steeply tilted,
  red spectrum, while the primordial gravitational waves have a flat
  spectrum. Instead, primordial density perturbations
  could be generated by a curvaton mechanism.
\end{abstract}

\maketitle

\section{Introduction}

Recent cosmological observations such as the WMAP results are
consistent with the inflationary scenario. Hence, we are prompted
to seek the inflaton in a unified theory of particle physics. 
Currently, it is widely believed that the most promising
candidate for a unified theory is superstring theory.
Interestingly, the superstring theory predicts the existence of the
extra dimensions. In order to reconcile this prediction
with our observed 4-dimensional universe, 
we need a mechanism to hide the extra dimensions.
For a long time, the Kaluza-Klein compactification was considered
to be the unique choice. However, recent developments
in superstring theory suggest a braneworld picture where we are living
on 4-dimensional hypersurface embedded 
in a higher dimensional spacetime~\cite{Horava}.
This braneworld picture not only gives a way for the superstring theory
to be phenomenologically viable but also suggests a new inflationary
scenario, so-called brane inflation~\cite{Dvali,Kachru}. 
 
In the brane inflation scenario, the radion, the distance between
branes, plays the role of the inflaton and inflation is terminated by
the brane collision.  This is nice because the inflation is realized
purely geometrical manner without introducing an ad-hoc scalar field.
In this scenario, however, branes are treated as test branes. 
On the other hand, relativistic cosmologists have studied the
braneworld gravity intensively~\cite{review}.  In these studies, the
effect of the bulk geometry on the 4-dimensional braneworld cosmology
have been a central concern, though inflation is usually assumed to be driven
by the fundamental scalar field either on the brane or in the bulk. 
In this braneworld cosmology, the self-gravity of branes are properly
treated and hence, it is clear how to calculate corrections due to the
bulk effect \cite{MWBH,LMWgw,KKS,HS}.
Taking a look at both approaches, we have come up with the idea of
incorporating geometrical inflation in a simple Randall-Sundrum (RS)
model~\cite{RS1}.  

In this paper, we would like to propose an inflationary scenario
driven by the flux generated by a brane that is charged with respect
to a five-form field strength. The idea is very
similar to that of brane inflation but we take into account the
self-gravity of branes. Our model is constructed in the RS
framework~\cite{RS1}.  We suppose that initially two positive tension
branes are inflating as de Sitter spacetimes embedded in an anti-de
Sitter bulk.  Eventually, they collide with each other and inflation
will end. Subsequently, two branes are assumed to coallesce and evolve
as a single $Z_2$-symmetric positive tension brane.  The gravitational
theory is non-singular and the model we construct is essentially
5-dimensional way.  Except at the collision point, we can use the
effective action obtained by the low energy 
approximation~\cite{KS1,KS2,wiseman}.  But in the
4-dimensional Einstein frame, the evolution of the universe is
discontinuous at the collision point, which means that the effective
4-dimensional theory breaks down.  However at the collision point we
can use 5-dimensional energy-momentum conservation to determine the
dynamics \cite{LMW}.  We are thus able to analyze the spectrum of
primordial scalar and tensor fluctuations produced after the collision. It
turns out that a curvaton-type mechanism is required to generate the
primordial density perturbations producing the present structure of
the universe.

The organization of this paper is as follows.
In sec.II, the basic setup is presented. In sec.III, our
cosmological scenario is described. Inflation driven 
by the flux is analyzed both in the induced metric frame and the
Einstein frame. The consistency analysis gives the
 expansion rate after the collision determined by the
 expansion rate of both branes before the collision. 
 The cosmological history after the collision is briefly
 summarized. 
 In sec.IV, the spectrum of fluctuations are calculated.
 The sec.V is devoted to conclusions. In the appendix, 
 the detailed derivation of the effective action is provided.   

\section{Basic Setup}

The point of brane inflation is that no fundamental scalar field is
necessary and the exit from inflation is realized by the collision of
branes. What we want to do is to incorporate this idea into the
codimension-one RS braneworld model.

We consider a two-brane system where one has a $Z_2$ symmetry 
and the other does not. Hereafter, we call the former the boundary brane
and the latter the bulk brane. 
As we show in an appendix such a set-up can be realised as the
limiting case of a three-brane system, with two boundary branes, where
one of the $Z_2$-symmetric branes is sent to infinity.

The model is described by the 5-dimensional action
\begin{eqnarray}
 S &=& \frac{1}{2\kappa^2} \int d^5 x \sqrt{-g}
                 \left[\overset{(5)}{R} + 2\Lambda \right]
 + \sum_{i=1}^2 \frac{1}{\kappa^2}  \int d^4 x \sqrt{-h_i} K^{i} \nonumber \\
&& -\frac{1}{2\cdot5!} \int d^5 x \sqrt{-g} F_5^2 
       + \sum_{i=1}^2 \mu_i \int C_4           \nonumber \\
&& + \frac{1}{4!} \int d^5 x \partial_A \left(\sqrt{-g} 
   F^{ABCDE} C_{BCDE} \right)    
   \nonumber \\
&&  + \sum_{i=1}^2  \int d^4 x \sqrt{-h_i } \left[-\sigma_i 
    + {\cal L}^i_{\rm matter} \right]
    \label{5d:action}  
\end{eqnarray}
where $\kappa^2$ is the 5-dimensional gravitational coupling constant
and $\overset{(5)}{R}$ is the 5-dimensional curvature.  Both our
branes have positive tension, $\sigma_1, \sigma_2>0$, but opposite
charges, $\mu_1>0,\ \mu_2<0$, which couple to a 4-form potential
field, $C_4 = (1/4!) C_{ABCD} dx^A \wedge dx^B \wedge dx^C \wedge dx^D
$.  The bulk brane separates the bulk into regions $I$ and $I\!I$ (see
FIG.1).

We denote the induced metric on each brane by $h_{i\mu\nu}$ and $K^i$
denotes the trace-part of the extrinsic curvature of each brane. Here
we have taken into account the Gibbons-Hawking boundary term instead
of introducing delta-function singularities in the five-dimensional
curvature.  We incorporated the 5-form field $F_5=dC_4$ which can change
the effective cosmological constant in the bulk $\Lambda$~\cite{TS}. The third
line represents the surface term which is introduced to make the
variation of the action with fixed $F_5$ consistent.

Let us take the coordinate system
\begin{eqnarray}
ds^2 = dy^2 + g_{\mu\nu}(y,x)dx^\mu dx^\nu 
\end{eqnarray}
The Latin indices $\{A,B,\cdots\}$ and the Greek indices $\{\mu,\nu,\cdots\}$ 
are used for tensors defined in the bulk and on the brane, respectively.  
The 5-form equation of motion becomes
\begin{eqnarray}
\partial_M\left(\sqrt{-g}F^{0123y} \right) dx^M
        +\sum_{i=1}^2 \mu_i\delta (y-\phi_i (x) ) dy =0
        \label{eom:5form}
\end{eqnarray}
where $\phi_i (x)$ denotes the position of each brane.
In each bulk ($y\neq \phi_i$), it is easy to solve Eq.~(\ref{eom:5form}) 
as
\begin{eqnarray}
  F_I^{0123y} = \frac{c_I}{\sqrt{-g}} \ , \quad
  F_{I\!I}^{0123y} = \frac{c_{I\!I}}{\sqrt{-g}} \ .
\end{eqnarray}
where $c_I, c_{II}$ are constants of integration. 
Because of the charge conservation, we have $c_{I\!I} =0$
 as is explained in the appendix.
Thus we see $F_5$ has no 
local dynamics. As $F_5$ is not dynamical, we can eliminate it from 
the action by simply substituting the above solution into the
original action. This can be done using the equations of motion
to give
\begin{eqnarray}
&& \frac{1}{4!} \int d^5 x \partial_A \left(\sqrt{-g} 
   F^{ABCDE} C_{BCDE} \right)  \nonumber\\
 &&  =  \frac{1}{5!} \int d^5 x F_5^2 
   + \int d^5 x \partial_y \left(\sqrt{-g} 
   F^{0123y}  \right)  C_{0123} \nonumber\\
 &&  = - \int_I d^5 x c_I^2 
   - \sum_{i=1}^2 \mu_i  \int d^4 x   C_{0123} \ .
 \label{eom2:5form}
\end{eqnarray}
Substituting Eq.~(\ref{eom2:5form}) into the original action
(\ref{5d:action}), we see that 4-form potentials are cancelled and the
effect of 5-form field strength is indistinguishable from a
cosmological constant term in the bulk. Thus, the resultant action is
\begin{eqnarray}
 S &=& \frac{1}{2\kappa^2} \int_I d^5 x \sqrt{-g}
        \left[\overset{(5)}{R} + 2\Lambda -\kappa^2 c_I^2 \right] \nonumber\\
  && + \frac{1}{2\kappa^2} \int_{I\!I} d^5 x \sqrt{-g}
        \left[\overset{(5)}{R} + 2\Lambda  \right] \nonumber\\    
 &&  + \sum_{i } \frac{1}{\kappa^2}  
           \int \sqrt{-h_i}K^{i} \nonumber \\
 &&  + \sum_i  \int d^4 x \sqrt{-h_i } \left[-\sigma_i 
     + {\cal L}^i_{\rm matter} \right]  \ .
\end{eqnarray}
The 5-form field strength $F_5$ in region $I$ acts like a change in
the effective 5-dimensional cosmological constant in region $I$,
$\Lambda_I\equiv \Lambda -\kappa^2 c_I^2/2$. Consequently, two bulk
regions have a different cosmological constant. The absolute value of
the cosmological constant in region I is assumed to be small and
therefore we shall see that the effective cosmological constants
induced on both branes are positive.  

If the expansion rate of the second brane is faster than that of
first, both branes will eventually collide with each other and the
opposing charges annihilate. The resulting brane tension is assumed to
become (close to) the Randall-Sundrum tuning value so that the
inflation ends and the universe becomes radiation dominated. We assume
a completely inelastic collision so that the kinetic energy of the
branes is transformed into radiation energy density on the brane. The
subsequent evolution is same as that of the radiation dominated
universe in the RSII brane model.  Thus, the original flux generated
by the charged branes has caused de Sitter inflation of branes and
the exit from inflation is realized by the brane collision.  In
the following sections, we shall look at the details of this scenario.

\begin{figure}[h]
\includegraphics[height=4cm, width=7cm]{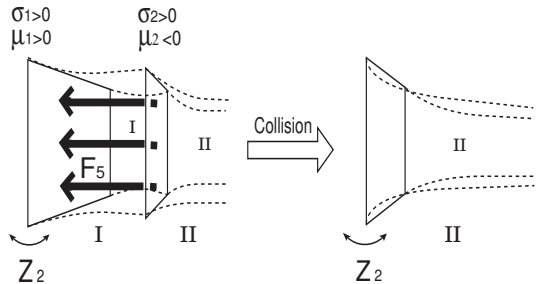}
\caption{The 3-brane charges produce $F_5$ flux in the region I.
This changes the cosmological constant in region I. After the collision,
the single brane remains and the big-bang universe commences.}
\label{fig:1}
\end{figure}
%

\section{Flux Inflation}

The non-linear dynamics of de Sitter branes embedded in 5-dimensional
anti-de Sitter space can be studied exactly without recourse to any
approximation \cite{kaloper,BCG,radion}. However the study of
inhomogeneous perturbations about this background, when two de Sitter
branes are in relative motion, is a much harder problem. Therefore we
will use a low-energy approximation \cite{KS1,wiseman,KS2,sugumi} in
this case, valid when the Hubble rate is smaller than the anti-de
Sitter scale. In this case the only extra degree of freedom coming
from the 5-dimensional gravitational field is the radion, a scalar in
the 4-dimensional effective theory, describing the distance between
the two branes.

Before the collision, the radion field is non-minimally coupled to the
4-dimensional gravitational field on either brane and the system
is described by the scalar-tensor theory. Hence, it is useful to look
at the cosmological evolution both from the induced metric frame on
the bulk brane and the Einstein frame.  After the collision the radion
field vanishes and the low-energy system is described by 4-dimensional
Einstein gravity.  As the collision changes the theory
discontinuously, the evolution of the universe in the Einstein frame
looks strange. We find a contracting universe immediately before the
collision which will start to expand abruptly after the collision.
However, in the induced metric frame which is a natural frame for an
observer, the universe is always expanding.

\subsection{Inflation in the induced metric frame}

Except for the collision point, the low energy approximation
can be applied~\cite{KS1,wiseman,KS2,sugumi}. 
The detailed derivation of the effective action for our system
can be found in the appendix. An alternative derivation is given in
Ref.~\cite{Ludo}. 
The induced metric on the bulk brane is $h_{\mu\nu}$.
The low-energy effective action on the bulk brane is
\begin{eqnarray}
S = \frac{\ell_I}{2\kappa^2} \int d^4 x \sqrt{-h} 
        \left[\left( \Psi^2 + \alpha -1 \right) R(h)  
        + 6(\partial \Psi )^2  - V  \right] \ ,
\end{eqnarray}
where $\Psi$ represents the radion field, with $\Psi\to1$ when the
branes are coincident. The AdS length scale
$\ell_I$ is given in Eq.~(\ref{eom:bulk}). For simplicity, we denote
$h^{\alpha\beta}\partial_\beta\Psi\partial_\alpha\Psi$ as
$(\partial\Psi)^2$.  The effective potential for the radion is given
by
\begin{eqnarray}
\frac{\ell_I^2}{12} V = \left(  \beta_1 -1 \right) \Psi^4  
         - \left( \frac{1}{\alpha} -1 - \beta_2 \right) \ .
\end{eqnarray}
Here we have defined the dimensionless parameters
\begin{eqnarray}
\alpha = \frac{\ell_{I\!I}}{\ell_I} \ , \quad
 \beta_1 = \frac{\kappa^2 \sigma_1 \ell_I}{6} \ , \quad
\beta_2 = \frac{\kappa^2 \sigma_2 \ell_I}{3} \ . 
\end{eqnarray}
We obtain the static case of single Minkowski brane at fixed distance
in AdS when $\beta_1=1$, $\beta_2=0$ and $\alpha=1$.
As we have $c_{I\!I}=0$, 
the effective cosmological constant in region $I$, 
$\Lambda_I=\Lambda-\kappa^2 c_I^2 /2 $, 
 is always smaller than that in region $I\!I$, $\Lambda$, 
 due to the flux in the region $I$. From (A2),
we see $\ell_I > \ell_{I\!I}$, i.e. $\alpha <1$.

\begin{figure}[h]
\includegraphics[height=5cm, width=7cm]{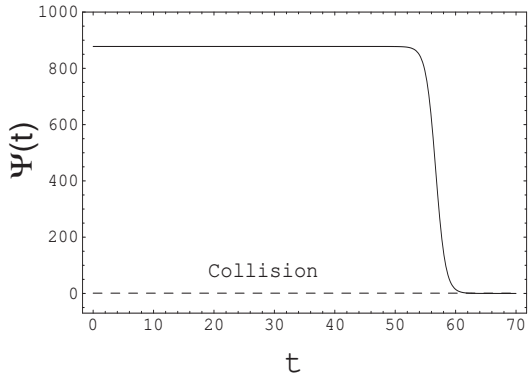}
\caption{Evolution of the radion in the induced metric frame is depicted for
 the parameters $\alpha=0.98 , \beta_1 = 1.000001, \beta_2 =0.005$. 
 For these parameters, Hubble 
parameter becomes 1.2413 in unit of $1/\ell_I$, 
the number of e-foldings becomes about 74.}
\label{fig:2}
\end{figure}

The equation of motion for the radion is
\begin{eqnarray}
\Box \Psi  - \frac{\Psi^2 -1 +\alpha }{12(1-\alpha )}
\frac{\partial V}{\partial \Psi} 
+ \frac{\Psi}{3(1- \alpha )} V(\Psi) = 0 \ .
\end{eqnarray}
The dynamics of the radion field thus appear non-trivial, and
as the effective theory in the induced metric frame is a
scalar-tensor theory, the cosmological dynamics will also
depend on this non-trivial dynamics. However,
the equations of motion for the induced metric can be
written as
\begin{eqnarray}
G_{\mu\nu}=- \frac{6}{\ell_I^2 (1-\alpha)} \left(
        \frac{1}{\alpha} -1 -\beta_2 \right) h_{\mu\nu}
        +E_{\mu\nu}
\end{eqnarray}
where $E_{\mu\nu}$ represents the projected 5-dimensional Weyl tensor
on the brane \cite{SMS} and is determined by the radion field,
\begin{eqnarray}
E_{\mu\nu}&=&
        \frac{6}{\ell_I^2(1-\alpha)} 
        \left(\frac{1}{\alpha} -1 -\beta_2 \right)h_{\mu\nu}
        \nonumber\\
&&      +\frac{2\Psi}{\Psi^2 -1 +\alpha } \left( \nabla_\mu \nabla_\nu \Psi
        -h_{\mu\nu} \Box \Psi \right) \nonumber\\
&&      -\frac{4}{\Psi^2 -1 +\alpha }
        \left( \nabla_\mu \Psi \nabla_\nu \Psi 
        -\frac{1}{4}h_{\mu\nu}(\partial\Psi)^2\right) \nonumber\\
&&      -\frac{1}{2(\Psi^2 -1 +\alpha) } h_{\mu\nu} V  \,.
\end{eqnarray}
This satisfies the traceless condition $E^\mu{}_\mu =0$ and indicates
the radion field behaves as the conformally
invariant matter on the brane. 
Hence, for the isotropic and homogeneous Universe,
\begin{eqnarray}
 ds^2 = -dt^2 + a^2 (t) \left( dx^2 + dy^2 + dz^2 \right) \ ,
\end{eqnarray}
the effect of the bulk gravity, $E_{\mu\nu}$, acts like a radiation fluid,
and the Friedman equation is obtained as
\begin{eqnarray}
  H_2^2 \ell_I^2 = \frac{2}{1-\alpha} \left(
      \frac{1}{\alpha} -1 -\beta_2 \right) + \frac{C}{a^4} \ ,
\end{eqnarray}
where $H_2$ denotes the Hubble parameter of the induced spacetime and 
the constant of integration $C$ is often called the dark radiation.
As we know $\alpha <1$, 
in order to have the positive effective cosmological constant,
 we assume 
\begin{eqnarray}
\frac{1}{\alpha} -1 -\beta_2 >0   \, .
\end{eqnarray}
In practice, once the universe starts to expand, the dark radiation term 
$\frac{C}{a^4}$ becomes soon negligible. Thus, 
we see that the induced spacetime on the brane rapidly
approaches de Sitter at late times ($a(t)\rightarrow\infty$).
We can obtain enough e-foldings on the brane, if we fine tune
the initial brane positions (and hence the initial value of $\Psi$,
see Fig.~\ref{fig:2}).
Thus, we have obtained an inflationary universe on the brane.

When $\Psi$ reaches $1$, the inflation is suddenly terminated
by the collision of branes in our scenario. 

\subsection{View from the Einstein frame}

We have shown that we can obtain a de Sitter inflationary universe
in the induced metric frame due to the existence of the flux between
the two branes in the bulk. It is interesting to see this 
in the conformally related Einstein frame~\cite{nojiri}.
In the Einstein frame, the metric is $\gamma_{\mu\nu} 
= [\Psi^2 + (\alpha -1)]~h_{\mu\nu}$, using the variable
$\Psi^2 = (1-\alpha) \coth^2 \psi $. Then the action reduces to
\begin{eqnarray}
S &=& \frac{\ell_I}{2\kappa^2} \int d^4 x \sqrt{-\gamma}
 \left[ R(\gamma)  - 6 \left( \partial \psi \right)^2   - U(\psi)
      \right] 
\end{eqnarray}
where the effective potential for the minimally-coupled radion $\psi$
is given by
\begin{eqnarray}      
\frac{\ell_I^2}{12}U=
        \left(\beta_1 -1 \right) \cosh^4 \psi   
        -\frac{ \frac{1}{\alpha} -1 - \beta_2}{(1-\alpha)^2}
        \sinh^4 \psi  \ .
\end{eqnarray}
\begin{figure}[h]
\includegraphics[height=5cm, width=7cm]{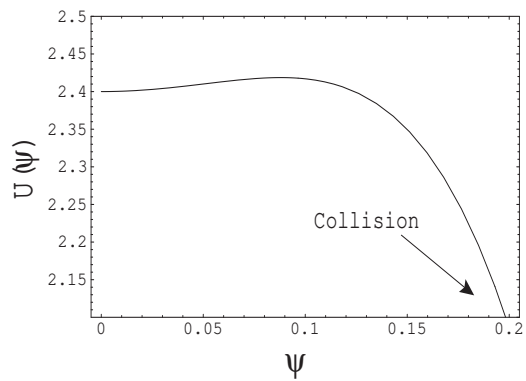}
\caption{Here, we present the radion potential. The collision point
is determined by $\coth^2 \psi = 1/\sqrt{1-\alpha}$.}
\label{fig:3}
\end{figure}
The above potential is depicted in Figure~\ref{fig:3}. 
The unstable extremum corresponds to the static two de Sitter brane
solution.  
\begin{figure}[h]
\includegraphics[height=5cm, width=7cm]{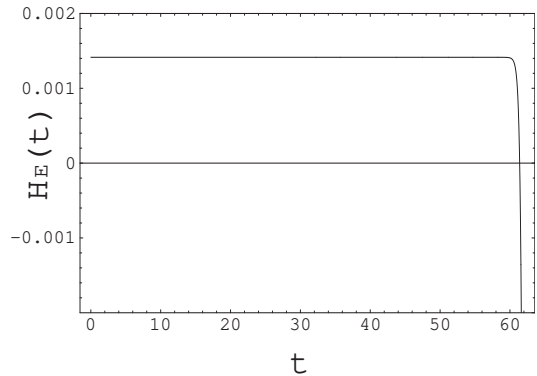}
\caption{Evolution of the Hubble parameter 
in the Einstein frame for the same parameters
 as Figure~\ref{fig:2}. The horizontal axis
 represents the cosmic time in Jordan frame. }
\label{fig:4}
\end{figure}
This extremum is located at $\psi_0$ determined by
\begin{eqnarray}
 (\beta_1 - 1) \cosh^2 \psi_0 
 = \frac{\frac{1}{\alpha}-1 -\beta_2}{(1-\alpha)^2} \sinh^2 \psi_0   \ .
\end{eqnarray}
In that case, the potential energy becomes
\begin{eqnarray}
\frac{\ell_I^2}{12} U \big|_{\psi =\psi_0}=
        \frac{\frac{1}{\alpha} -1 -\beta_2 }
        {(1-\alpha )^2 } \sinh^2 \psi_0  \ .
\end{eqnarray}
The effective mass-squared is negative,
\begin{eqnarray}
\frac{\ell_I^2}{12}\frac{d^2U}{d\psi^2} \Big|_{\psi =\psi_0}
        =-8 \frac{\frac{1}{\alpha} -1 -\beta_2 }
        {(1 - \alpha )^2 } \sinh^2 \psi_0 \ .
\end{eqnarray}
From the above, one can read off the radion effective mass $m^2_{r} =
-4 H_0^2 $, where $H_0$ represents the Hubble parameter at $\psi_0$.
This indicates the linearised instability of the static two de Sitter brane
system,
consistent with previous analyses \cite{GenSas,CF,Contaldi}.

Let us assume two branes are separated enough initially, which means
the radion is located at near the maximum.  The radion then starts to
roll down the hill and reaches the collision point $\coth^2 \psi =
1/\sqrt{1-\alpha}$ which corresponds to $\Psi=1$.  In contrast to the
other models discussing
collisions~\cite{Khoury,Gen,Bucher,KSS,GT,Pillado}, our model has no
singularity at the collision because the bulk region never disappears
in our model.

The Hubble parameter in the Einstein frame, $H_E$, is related to the 
Hubble parameter in the induced metric frame, $H_2$, as
\begin{eqnarray}
  H_E = \frac{1}{\sqrt{\Psi^2 -1 +\alpha}} H_2 
        + \frac{\Psi\dot{\Psi}}{\left(\Psi^2 -1 + \alpha \right)^{3/2}} \ .
\end{eqnarray}
The typical behavior of the Hubble parameter in the Einstein frame
is depicted in Figure~\ref{fig:4}. 
While $\Psi$ remains almost
constant the Hubble rate is also nearly constant and we have almost
exponential expansion in both the induced and Einstein frames.
Due to the suppression factor $1/\sqrt{\Psi^2 -1 +\alpha}$, the Hubble
rate in the Einstein frame is much smaller than that in the induced
frame. However, the e-folding number is almost the same as that in the
Jordan frame because the time in the Einstein frame becomes longer by
the conformal factor $\sqrt{\Psi^2+(\alpha-1)}$.  Shortly before the
collision, we find that the effective potential in the Einstein frame
becomes negative, the universe recollapses and immediately before the
collision the universe is contracting in the Einstein frame, though
the universe is always expanding in the induced metric frame.

\subsection{Graceful Exit Through Brane Collision}

We need a fully 5-dimensional consideration to give a rule for
evolution through the collision. We consider the simplest case of a
completely inelastic collision where the bulk brane is absorbed by the
boundary brane.

\begin{figure}[h]
\includegraphics[height=5.5cm, width=5.5cm]{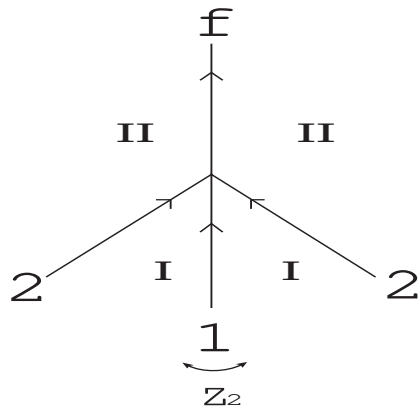}
\caption{The collision process is schematically depicted.
The final brane has also $Z_2$ symmetry.}
\label{fig:5}
\end{figure}

To describe the $Z_2$-symmetric collision of the branes it is useful
to consider the complete system of four branes, consisting of the
incoming $Z_2$ symmetric boundary brane, two copies of the bulk brane,
and the outgoing $Z_2$ symmetric brane (see Figure~\ref{fig:5}).

A detailed analysis gives consistency conditions for the
collision~\cite{Neronov,LMW}. These are equivalent to relativistic
energy-momentum conservation~\cite{LMW}
\begin{eqnarray}
\label{rhofhi}
\rho_f=\sigma_1+2 \gamma_{2|1} \sigma_2
\end{eqnarray}
where the pseudo-Lorentz factor between the colliding branes can be
written as $\gamma_{2|1} \equiv \cosh(\theta_2-\theta_1)$ where
\begin{equation}
\sinh\theta_1 = H_1\ell_1 \,, \quad
\sinh\theta_2 = H_2\ell_1 \,.
\end{equation}

At low energy (small ``angles'') this reduces to the ``Newtonian''
energy conservation law 
\begin{eqnarray}
\label{rhoflo}
\rho_f=\sigma_1+2 \sigma_2 \,,
\end{eqnarray}
plus the momentum conservation law
\begin{eqnarray}
H_f=\frac{1}{\alpha} \left[ H_1 - (1-\alpha )H_2 \right] \ .
\end{eqnarray}
Hence, in order to obtain an expanding universe after the collision
we have to impose the constraint
\begin{eqnarray}
  H_1 > (1-\alpha ) H_2 \, .
\end{eqnarray}
In addition to this constraint, we require that the expansion rate of
the bulk brane is bigger than the $Z_2$ symmetric brane, $H_2 > H_1$,
in order to cause a collision. It is easy to meet both these requirements,
as we set $\alpha <1$.

\subsection{Cosmological Evolution After the Collision}

After the collision, we assume the branes coalesce and behave as a
single brane.  We further assume the resulting brane tension is given
by the RS value, $\kappa^2\sigma_f=6/\ell_{II}$ so that the effective
cosmological constant on the brane vanishes after the collision. This
is the brane-world equivalent of the usual assumption that the
inflaton potential is zero at its minimum.

At the collision, the additional energy density (above the RS brane
tension) is assumed to be transferred to light degrees of freedom on
the brane, i.e., radiation.  Hence, the subsequent evolution will be
governed by the standard Friedmann equation for the hot big bang with
small Kaluza-Klein corrections.
 
As the radion disappears after the collision, the difference between
the induced metric frame and the Einstein frame also disappears.  In
the Einstein frame, the abrupt change of the contracting phase to the
expanding phase cannot be described within the 4-dimensional effective
theory. This clearly shows that our model is different from a
conventional 4-dimensional inflationary model.


\section{Perturbations}

Having constructed a homogeneous cosmological model, we can now
consider the spectrum of inhomogeneous perturbations that would be
expected due to small-scale quantum fluctuations.  As there is no
singularity at the collision, we can unambiguously follow the
evolution of fluctuations generated during inflation through the
collision.

To study the behavior of  fluctuations before the collision,
it is  convenient to work in the Einstein frame in which the radion is
minimally coupled to the metric. 
We have the equations of motion in the Einstein frame
\begin{eqnarray}
&&G_{\mu\nu}=6 \left( \partial_\mu \psi \partial_\nu \psi
        -\frac{1}{2}\gamma_{\mu\nu}(\partial\psi)^2\right) 
        -\frac{1}{2}\gamma_{\mu\nu } U (\psi)  \nonumber \\
&&      \Box\psi 
        -\frac{1}{12} \frac{d U}{d \psi} =0 \ .
\end{eqnarray}
The homogeneous and istropic background metric is
\begin{eqnarray}
ds^2 = b^2 (\eta ) \left[ - d\eta^2 + \delta_{ij} dx^i dx^j \right] \ .
\end{eqnarray}
and we have the Einstein equations become
\begin{eqnarray}
&&3{\cal H}^2 
        =3\psi'^2+\frac{1}{2}b^2U(\psi)\\
&&{\cal H}^2+2{\cal H}'
        =-3\psi'^2+\frac{1}{2}b^2U(\psi )    \\
&&\psi''+2{\cal H}\psi' 
        +\frac{ b^2}{12}\frac{\partial U}{\partial \psi}=0 \ .
\end{eqnarray}
where a prime denotes a derivative with respect to the conformal
time $\eta$ and ${\cal H}=b'/b$.
During the de Sitter phase, the solution is
\begin{eqnarray}
b=\frac{1}{- H_0 \eta}  \ , \quad
\psi = \psi_0 \ ,
\label{deSb}
\end{eqnarray}
where $bH_0={\cal H}_0$ and
\begin{eqnarray}
\ell_I^2 H_0^2 = 2\frac{\frac{1}{\alpha}-1-\beta_2}{(1-\alpha)^2}
            \sinh^2 \psi_0  \ .
\end{eqnarray}
Now we can examine possible fluctuations separately.

\subsection{Gravitational waves}

Let us consider first the tensor perturbations
\begin{eqnarray}
ds^2=b^2 (\eta ) \left[ -d\eta^2 
        +\left( \delta_{ij} + q_{ij} \right) dx^i dx^j \right] \ ,
\end{eqnarray}
where the tensor perturbations satisfy 
$q^i{}_i =0 \ , \ q_{ij,j}=0$. We can reduce the Einstein equations
to
\begin{eqnarray}
q_{ij}'' + 2 {\cal H} q_{ij}' + k^2 q_{ij} = 0 \ .
\end{eqnarray}
Before the radion starts to roll down the hill, the background
spacetime is the de Sitter spacetime (\ref{deSb}).
The positive frequency mode function is the standard one
\begin{eqnarray}
  q_{ij} \propto (- H_0 \eta)^{3/2} H^{(1)}_{3/2} ( -k \eta ) \ ,
\label{qoft}
\end{eqnarray}
where $H^{(1)}_{3/2}$ is the Hankel's function of the first kind.
This gives the standard flat spectrum for the primordial gravitational
waves. During the roll down phase, the universe will begin to contract
rapidly from a numerical calculation, we see that the contracting
phase is negligible in practice. During the collision process, as the
gravitational waves are independent of gauge, we will have a flat
spectrum on long wavelengths after the collision when the standard radiation
dominated era begins.

\subsection{Radion fluctuations}

To study the behavior of the radion fluctuations,
we express the metric perturbation in the Einstein frame as
\begin{eqnarray}
&&ds^2=b^2\left[
        -(1+2A)d\eta^2+2\partial_iBdx^id\eta\right.
        \nonumber\\
        &&\qquad\qquad\left.
        +\left((1+2{\cal R})\delta_{ij}+2\partial_i\partial_jE\right)
        dx^idx^j\right] \ ,
        \label{mtrc:sclr-ptb}
\end{eqnarray}
where $A, B, {\cal R}, E$ represent the gauge-dependent scalar metric
perturbations. A convenient gauge-invariant combination is the
comoving curvature perturbation, which is the intrinsic curvature
perturbation on uniform-radion hypersurfaces:
\begin{eqnarray}
{\cal R}_c={\cal R}-{\cal H}{\delta\psi\over\psi'}\,.
\end{eqnarray}

The second-order action for the curvature perturbation ${\cal R}_c$ is
\begin{eqnarray}
S={1\over2}\int d\eta\, d^3x\,z^2
\left[{\cal R}_c'{}^2-{\cal R}_c^{\,|i}{\cal R}_{c\,|i}\right]\,,
\label{ptb:action}
\end{eqnarray}
where
\begin{eqnarray}
z=\sqrt{3\ell_I \over2\kappa^2}\,{b\psi'\over{\cal H}}\,.
\label{Rcdef}
\end{eqnarray}
The equation of motion for ${\cal R}_c$ is
\begin{equation}
      {\cal R}_c''+ 2{z'\over z}{\cal R}_c' + k^2{\cal R}_c = 0 \ .
\label{sclr:schrodinger-typ}
\end{equation}
Therefore, on large scales, ${\cal R}_c$ is constant.

Equivalently we can work in terms of the radion on uniform-curvature
hypersurfaces
\begin{eqnarray}
\label{QR}
Q = \delta\psi - \frac{\psi'}{{\cal H}} {\cal R} = -
\frac{\psi'}{{\cal H}} {\cal R}_c \,.
\end{eqnarray}
During the de Sitter phase,  
the equation for $Q$ can be written as
\begin{eqnarray}
Q'' + 2 {\cal H} Q' + (k^2- 4H_0^2 b^2) Q =0 \ ,
\end{eqnarray}
where we have used Eq.~(31). 
Hence, the positive frequency mode becomes
\begin{eqnarray}
Q \sim (-H_0 \eta)^{3/2} H^{(1)}_{5/2} (-k\eta ) \ ,
\end{eqnarray}
where $H^{(1)}_{5/2} $ is the Hankel's function of the first kind.
Thus, we can read off the power spectrum of $Q$ as ${\cal
  P}_{Q} \sim k^{-2}$. Despite the exponential expansion the
spectrum on large scales becomes red during the de Sitter inflation
because the radion has negative effective mass-squared. 

These field fluctuations can be translated to the curvature perturbations on
comoving hypersurfaces via Eq.~(\ref{QR}). The coefficient
$\psi'/{\cal H}$ is independent of scale and hence the comoving
curvature perturbation shares the same red spectrum, ${\cal
  P}_{\cal R} \sim k^{-2}$.
Near the collision, we expect the spectrum to be blue because of the rapid
contraction, but this only affects small scales.

Finally, we need to calculate the curvature perturbation on
uniform-density hypersurfaces, $\zeta$, on large scales after the
collision. This should then remain constant on large scales for
adiabatic perturbations after the collision, even in the
brane-world~\cite{Langlois}, simply as a consequence of local energy
conservation \cite{WMLL}. 

In the low-energy limit, energy conservation at the collision gives
Eq.~(\ref{rhoflo}) which implies that the collision hypersurface will
be a uniform-energy hypersurface. Thus $\zeta$ after the collision
coincides with the curvature perturbation of this collision
hypersurface.

We define the collision hypersurface in terms of the low-energy
effective theory before the collision by the condition that
$\Psi=1$. Thus the collision hypersurface is a 
uniform-radion hypersurface and we can identify the curvature
perturbation on this hypersurface as the comoving curvature
perturbation, $\zeta=-{\cal R}_c$ (where the negative sign comes from
different historical conventions for the sign of the curvature
perturbation).

One might worry that ${\cal R}_c$ was calculated in the Einstein frame
and the collision hypersurface corresponds to a physical hypersurface
on the two branes. In fact the comoving curvature perturbation is
invariant under any conformal transformation that is function of the
radion field, $\Omega^2(\Psi)$, as the conformal transformation then
corresponds to a uniform rescaling on uniform-$\Psi$ hypersurfaces,
leaving the comoving curvature perturbation conformally
invariant. Hence ${\cal R}_c$ does describe the curvature perturbation
on the physical collision hypersurface.

Alternatively we can calculate the curvature perturbation on the
collision hypersurface using the $\delta N$
formalism~\cite{Sasaki}. This says that the 
curvature perturbation is given by the
perturbed expansion with respect to an initial flat hypersurface, or
$\zeta = (dN/d\phi_*) \delta\phi_* $
where $\delta\phi_* $ is the field perturbation on the initial spatially flat
hypersurface at horizon exit. This gives the usual expression
${\cal R}_c = - (H/\dot\psi) Q $ for the comoving curvature
perturbation during inflation.
Because the collision hypersurface is a uniform-$\psi$ hypersurface
before the collision and a uniform-density hypersurface after the
collision we have $\zeta=-{\cal R}_c = (H/\dot\psi) Q$.

However, we have shown that the comoving curvature perturbation has a
steeply tilted red spectrum which means that these perturbations
cannot be responsible for the formation of the large scale structure
in our Universe.

\subsection{Scale-invariant perturbations}

The curvaton mechanism \cite{EnqSlo,LW,MT} provides a simple example
of how scale-invariant perturbations could be generated in our model.
Both the bulk and boundary branes experience a de Sitter expansion
before the collision. The radion field acquires a red spectrum due to
its negative mass-squared, but any minimally-coupled, light scalar
field living on either brane would acquire a scale-invariant spectrum
of perturbations before the collision.

Thus we add to our basic model an additional degree of freedom which,
to be specific, we assume lives on the bulk brane.
The action for the curvaton $\chi$ in the induced metric frame 
takes the simple form
\begin{eqnarray}
S_{\rm curvaton}=\int d^4 x \sqrt{-h} \left[ -\frac{1}{2} (\partial \chi )^2
 \right]
\end{eqnarray}
Vacuum fluctuations of a massless scalar field in de Sitter
spacetime have the same time-dependence as the graviton (\ref{qoft}),
becoming frozen-in on large scales.
Normalising to the Bunch-Davies vacuum state on small scales leads to
the standard scale-invariant spectrum, $(H_2/2\pi)^2$, on super-Hubble
scales. Hence, the fluctuations of the curvaton have a completely flat
spectrum.

This is similar the way the curvaton mechanism was originally proposed
in the pre-big bang scenario \cite{EnqSlo,LW} where axion fields have a
non-trivial coupling to the dilaton field in the Einstein frame, but
are minimally coupled in a conformally related frame, which can give
rise to a scale-invariant spectrum if the expansion is de Sitter in
that conformal frame \cite{CEW}.

The curvaton mechanism requires that the curvaton has a non-zero mass
after the collision, which seems entirely natural if the collision is
associated with some symmetry-breaking in the degrees of freedom on
the brane. The energy density of any massive scalar field will grow
relative to the radiation density once the Hubble rate drops below its
effective mass. Indeed the particles must decay before primordial
nucleosynthesis to avoid an early matter domination in conflict with
standard predictions of the abundances of the light elements.  But if
the curvaton comes to contribute a significant fraction of the total
energy density before it decays then perturbations in the curvaton
will result in primordial density perturbations on large scales
capable of seeding the large scale structure in our Universe
\cite{EnqSlo,LW,MT}.

\section{Conclusions}

We have proposed a geometrical model of inflation where inflation on
two branes is driven by the flux generated by the brane charge and
terminated by the brane collision with charge annihilation.
We assume the collision process is completely inelastic and the kinetic
energy is transformed into the thermal energy. 
After the collision the two branes coalesce together
and behave as a single brane universe with zero effective cosmological
constant.  
 
The four-dimensional, low-energy effective theory in the Einstein
frame has to change abruptly at the collision point. Therefore, our
model needs to be consistently described using 5-dimensional gravity.
This can be done using consistency conditions at the collision that
ensure energy-momentum conservation in the 5-dimensional theory.  As
the collision process has no singularity in the 5-dimensional gravity,
we can unambiguously follow the evolution of inhomogeneous vacuum
fluctuations about this homogenous background during the whole history
of the universe. It turns out that the radion fluctuations produced
during inflation have a steep red spectrum and cannot produce the
present large-scale structure of our universe.  Instead the curvature
perturbations observed today must be generated by the curvaton or some
similar mechanism from initially isocurvature excitations of massless
degrees of freedom on one or other of the branes before the collision.
The primordial gravitational waves produced are likely to be difficult
to detect.

Our model does suffer from a significant fine tuning problem of the
initial conditions. The radion field must start very close to an
unstable extremum of its potential in order to obtain sufficient
inflation.  However, the radion field has a geometrical meaning so
that the initial value is determined by the initial separation between
two branes.  Sufficient separation of the initial brane positions does
not seem so unnatural, and the fine tuning valu of the radion may not
be so serious.  A quantum cosmological consideration~\cite{GS,Koyama}
would be needed to give a more insight on this initial condition problem.
It might be that one can find an instanton solution to describe
tunnelling to this extremum. Alternatively, in a semi-classical model
such as stochastic inflation, it may be that quantum fluctuations
drive the field to the top of the potential before classical evolution
takes over and the field rolls down, leading to the branes to collide.

We stress that our model is fundamentally 5-dimensional in nature.  It
cannot be constructed starting from a 4-dimensional theory. It would
of course be interesting to see if it is possible to embed our model
into string theory model, but that goes beyond the scope of this
paper.

{\em Note added:} While this work was being written up Koyama and Koyama
\cite{KKKK} submitted a related paper deriving an equivalent effective
action for anti-D-branes in a type IIB string model.

\acknowledgments 
We wish to thank Roy Maartens for useful discussions. 
 SK and JS would like to thank 
 the Portsmouth ICG for its hospitality
  and financial support, under PPARC grant PPA/V/S/2001/00544.
  This work is supported by the
  Grant-in-Aid for the 21st Century COE "Center for Diversity and
  Universality in Physics" from the Ministry of Education, Culture,
  Sports, Science and Technology (MEXT) of Japan.  This work was also
  supported in part by Grant-in-Aid for Scientific Research Fund of
  the Ministry of Education, Science and Culture of Japan No. 155476
  (SK), No.14540258  and No.17340075 (JS).

\appendix

\section{Derivation of effective action}

\subsection{Static Solution}

We shall start with the three-brane system and take the
two-brane limit to derive the low energy
effective action for the two-brane system.
This allows us to use a simple moduli approximation method.
Assuming there exists no matter on the third brane, this
procedure can be justified. 

In the low energy limit, the configuration should be almost static.
As we have no matter in the bulk, the bulk metric should be
anti-de Sitter spacetime. 
Hence, let us first consider the static solution of the form
\begin{eqnarray}
&& ds_I^2 = dy^2 + b_I^2 (y) \eta_{\mu\nu} dx^\mu dx^\nu \nonumber\\
&& ds_{I\!I}^2 = dy^2 + b_{I\!I}^2 (y) \eta_{\mu\nu} dx^\mu dx^\nu
\end{eqnarray}
where $b_I = \exp (- y/\ell_I )$ and 
$b_{I\!I} = \exp (- y/\ell_{I\!I} )$ represents the warp factor in each region. 
The bulk equation of motion imply the relation 
between the curvature scale in the bulk 
and the flux as
\begin{eqnarray}
&&  \frac{6}{\ell_I^2} = \Lambda -\frac{\kappa^2}{2} c_I^2 \ , \nonumber\\
&&  \frac{6}{\ell_{I\!I}^2} = \Lambda -\frac{\kappa^2}{2} c_{I\!I}^2  \ .
\label{eom:bulk}
\end{eqnarray}
We also have junction conditions for the metric which give the relation
\begin{eqnarray}
 && \kappa^2 \sigma_1 = \frac{6}{\ell_I} \\
 && \kappa^2 \sigma_2 = \frac{3}{\ell_{I\!I}} -\frac{3}{\ell_I} \\
 && \kappa^2 \sigma_3 = -\frac{6}{\ell_{I\!I}} \ .
\end{eqnarray}
If we assume the inflating bulk brane $\sigma_2 >0$, we have 
$\ell_{I\!I} < \ell_I $.
The above implies the relation
\begin{eqnarray}
   \sigma_1 + 2\sigma_2 + \sigma_3 =0 \ .
   \label{econservation}
\end{eqnarray}
The junction conditions for 5-form fields give
\begin{eqnarray}
 && c_I =- \frac{\mu_1}{2}  \ , \quad 
c_{I\!I} = -(\mu_2 +\frac{\mu_1}{2}) \\ 
 && \mu_1 + 2\mu_2 + \mu_3 =0  \ . 
 \label{chgconservation}
\end{eqnarray}
The last relation is nothing but the charge conservation law. 
The static solution is realized under the relations Eqs.~(\ref{econservation}) 
and (\ref{chgconservation}).
In this paper, we set $\mu_3 =0$. This implies $c_{I\!I} =0 $.

\subsection{Low energy effective action}

\begin{figure}[h]
\vspace{-5mm}\hspace{-1cm}
\includegraphics[height=4cm, width=5cm]{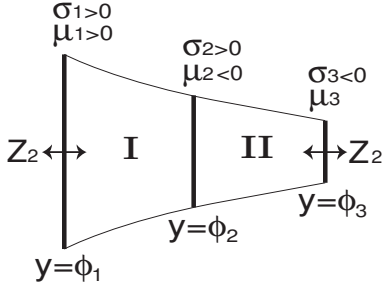}
\caption{three-brane system}
\end{figure}

Let us employ the moduli approximation method~\cite{GPT} 
which can be shown to be valid at low energy~\cite{Soda,Kanno}.
(An alternative derivation appears in Ref.~\cite{Ludo}.)
In this method, we can use the factorized metric ansatz
\begin{eqnarray}
ds^2 &=& dy^2 + b_I^2 (y) g_{\mu\nu}^I (x) dx^\mu dx^\nu  \nonumber\\
ds^2 &=& dy^2 + b_{I\!I}^2 (y) g_{\mu\nu}^{I\!I} (x) dx^\mu dx^\nu  \ . 
\label{fmetric}
\end{eqnarray}
The brane positions are denoted by $\phi_1 (x), \phi_2 (x)$ 
and $\phi_3 (x)$. 
Now, we calculate the bulk action, the action for each brane tensions,
and the Gibbons-Hawking terms, separately. 
 
Let us start with the calculation of the bulk action.
For the metric (\ref{fmetric}), using the Gauss equation,
it is straightforward to relate the bulk Ricci scalar to the
4-dimentional Ricci scalar, 
\begin{eqnarray}
   \overset{(5)}{R} = \frac{R(g^i)}{b_i^2} - \frac{20}{\ell_i^2}   \ ,
\end{eqnarray}
we can write the contributions from each bulk as
\begin{eqnarray}
S_{\rm bulk I}&=&\frac{2}{2\kappa^2} \int d^4 x \sqrt{-g^I}
        \int^{\phi_2}_{\phi_1}dy b_I^4(y)\left[
        \frac{R(g^I)}{b_{I}^2} - \frac{8}{\ell_{I}^2} \right] \nonumber\\
&=&     \frac{1}{\kappa^2} \int d^4 x \sqrt{-g^I} \left[
        -\frac{\ell_I}{2} \left\{ b_I^2 (\phi_2 )
        -b_I^2(\phi_1 )\right\}R(g^I)\nonumber \right. \\
&&      \left. +\frac{2}{\ell_I} \left\{ b_I^4 (\phi_2 )
        - b_I^4 (\phi_1 )  \right\}\right]
        \label{bulkI}
\end{eqnarray}
where we used the bulk equation of motion (\ref{eom:bulk}) in the
first line. The factor 2 over $\kappa^2$ comes from the $Z_2$ symmetry 
of this spacetime and we neglected the second order quantities.
In the same way for region II, we have
\begin{eqnarray}
&&\hspace{-5mm}
S_{\rm bulk I\!I}=\frac{2}{2\kappa^2} \int d^4 x \sqrt{-g^{I\!I}}
        \int^{\phi_3}_{\phi_2} dy b_{I\!I}^4 (y) \left[
        \frac{R(g^{II})}{b_{I\!I}^2}-\frac{8}{\ell_{I\!I}^2} \right]
        \nonumber\\
&&      =\frac{1}{\kappa^2} \int d^4 x \sqrt{-g^{I\!I}} \left[
        -\frac{\ell_{I\!I}}{2} \left\{ b_{I\!I}^2 (\phi_3 )
        -b_{I\!I}^2 (\phi_2 )\right\}R(g^{II})\nonumber \right. \\
&&      \left.\qquad
        +\frac{2}{\ell_{I\!I}} \left\{ b_{I\!I}^4 (\phi_3 )
        -b_{I\!I}^4 (\phi_2 )\right\}\right]\ .
        \label{bulkII}
\end{eqnarray}

In order to calculate the action for each brane tensions,
we need to know the induced metric. 
The induced metric on the left boundary brane becomes
\begin{eqnarray}
   b^2 (\phi_1 ) g_{\mu\nu}^I 
              + \partial_\mu \phi_1 \partial_\nu \phi_1
\end{eqnarray}
and the similar formula holds for other branes. 
Then, the contributions from the brane tension terms become
\begin{eqnarray}
S_1 = - \sigma_1 \int d^4 x \sqrt{-g^I} b_I^4 (\phi_1 )
  \left[ 1+ \frac{1}{2 b_I^2 (\phi_1 ) } \left( 
  \partial \phi_1 \right)^2 \right] \ ,\quad
  \label{tension1}
\end{eqnarray}
\begin{eqnarray}
S_2&=&-2\sigma_2 \int d^4 x \sqrt{-g^I} b_I^4 (\phi_2 )
        \left[ 1+ \frac{1}{2 b_I^2 (\phi_2 ) } \left( 
        \partial \phi_2 \right)^2 \right] \nonumber\\
&=& \!\!\!\!
        -2\sigma_2 \int d^4 x \sqrt{-g^{I\!I}} b_{I\!I}^4 (\phi_2 )
        \left[ 1+ \frac{1}{2 b_{I\!I}^2 (\phi_2 ) } \left( 
        \partial \phi_2 \right)^2 \right] \qquad\quad
        \label{tension2}
\end{eqnarray}
and
\begin{eqnarray}
S_3 = - \sigma_3 \int d^4 x \sqrt{-g^{I\!I}} b_{I\!I}^4 (\phi_3 )
  \left[ 1+ \frac{1}{2 b_{I\!I}^2 (\phi_3 ) } \left( 
  \partial \phi_3 \right)^2 \right]   \ , \quad
  \label{tension3}
\end{eqnarray}
where the factor 2 in front of $\sigma_2$ comes from the $Z_2$
symmetry. 

Let us turn to the calculation of Gibbons-Hawking terms.
The extrinsic curvature in leading order is given by
\begin{eqnarray}
 K_{\mu\nu} = n_y
 \left[ \frac{4}{\ell} +\frac{1}{b^2}\Box \phi
 + \frac{1}{\ell b^2 } (\partial \phi )^2 \right] \ ,
\end{eqnarray}
where $n_y$ is the normal vector to the brane defined by 
\begin{eqnarray}
  n_y = \left( 1+ \frac{1}{b^2} (\partial \phi )^2 \right)^{-1/2} \ .
\end{eqnarray}
Hence, the Gibbons-Hawking terms are
\begin{eqnarray}
S_{\scriptscriptstyle \rm GH1}
=\frac{2}{\kappa^2 \ell_I} \int d^4 x \sqrt{-g^I} \left[
        4 b_I^4 (\phi_1 ) + 3 
        b_I^2(\phi_1 ) \left( \partial \phi_1 \right)^2 \right] \ ,\ \quad
        \label{GH1}
\end{eqnarray}
\begin{eqnarray}
S_{\scriptscriptstyle\rm GH2}
=\frac{2}{\kappa^2 \ell_{I\!I}} \int d^4 x \sqrt{-g^{I\!I}} \left[
        4 b_{I\!I}^4 (\phi_2 ) + 3 
        b_{I\!I}^2 (\phi_2 ) \left( \partial \phi_2 \right)^2 \right] 
        \nonumber\\
        -\frac{2}{\kappa^2 \ell_I} \int d^4 x \sqrt{-g^I} \left[
        4 b_I^4 (\phi_2 ) + 3 
        b_I^2 (\phi_2 ) \left( \partial \phi_2 \right)^2 \right] \ ,
        \qquad
        \label{GH2}
\end{eqnarray}
 and
\begin{eqnarray}
S_{\scriptscriptstyle \rm GH3}
=\frac{2}{\kappa^2 \ell_{I\!I}} \int d^4 x \sqrt{-g^{I\!I}} \left[
        4 b_{I\!I}^4 (\phi_3 ) + 3 
        b_{I\!I}^2 (\phi_3 ) \left( \partial \phi_3 \right)^2 \right] \ ,\ \ 
        \label{GH3}
\end{eqnarray}
where the factor 2 over $\kappa^2$ in $S_{\rm GH2}$ again 
comes from the $Z_2$ symmetry

Substituting Eqs.(\ref{bulkI}), (\ref{bulkII}), (\ref{tension1}), 
(\ref{tension2}) (\ref{tension3}), (\ref{GH1}), (\ref{GH2}) and (\ref{GH3})
into the 5-dimensional action (\ref{5d:action}), the resultant 4-dimensional
effective action can be summarized by the following.
The curvature part becomes
\begin{eqnarray}
&&\hspace{-8mm}S_{\rm R}=\frac{\ell_I}{2\kappa^2} \int d^4 x \sqrt{-g^I} \left[
     b_{I}^2 (\phi_1 ) -  b_I^2 (\phi_2 ) \right] R (g^I) \nonumber\\
&&\hspace{-5mm}
        + \frac{\ell_{I\!I}}{2\kappa^2} \int d^4 x \sqrt{-g^{I\!I}} \left[
     b_{I\!I}^2 (\phi_2 ) 
     -  b_{I\!I}^2 (\phi_3 ) \right] R (g^{I\!I}) \ . \quad
\end{eqnarray}
The kinetic part of radions are
\begin{eqnarray}
&&\hspace{-5mm}
S_{\rm K}=\frac{3}{\kappa^2 \ell_I} \int d^4 x \sqrt{-g^I} \left[
       b_I^2 (\phi_1 ) \left( \partial \phi_1 \right)^2
     - b_I^2 (\phi_2 ) \left( \partial \phi_2 \right)^2 \right] \nonumber\\
&&\hspace{-2mm}
     + \frac{3}{\kappa^2 \ell_{I\!I}} \int d^4 x \sqrt{-g^{I\!I}} \left[
       b_{I\!I}^2 (\phi_2 ) \left( \partial \phi_2 \right)^2
     - b_{I\!I}^2 (\phi_3 ) \left( \partial \phi_3 \right)^2 \right] \ . \qquad
\end{eqnarray}
The potential energy will be induced as
\begin{eqnarray}
&&\hspace{-1.7cm}S_{\rm V}=\int d^4 x \sqrt{-g^I } \left[
        \left(\frac{6}{\kappa^2 \ell_I } - \sigma_1 \right) b_I^4 (\phi_1 )
        \nonumber \right.\\
&&\hspace{-5mm}\left.
         - \left\{ 2\sigma_2 -\frac{6}{\kappa^2}\left(
         \frac{1}{\ell_{I\!I}} - \frac{1}{\ell_{I}} \right) \right\} 
         b_I^4 (\phi_2 ) \right]  \nonumber\\ 
&&\hspace{-5mm}
         -\int d^4 x \sqrt{-g^{I\!I}} 
        \left( \sigma_3 + \frac{6}{\kappa^2 \ell_{I\!I}} \right) 
        b_{I\!I}^4 (\phi_3 ) \ .
\end{eqnarray}

Now we can write down the effective action for the bulk brane.
The continuity condition of the metric is given by
\begin{eqnarray}
  h_{\mu\nu} = b_{I}^2 (\phi_2 ) g^I_{\mu\nu} 
             = b_{I\!I}^2 (\phi_2 ) g^{I\!I}_{\mu\nu} \ ,
\end{eqnarray}
where $h_{\mu\nu}$ is the induced metric of the bulk brane.
Defining the variables
\begin{eqnarray}
 \Psi = \frac{b_{I} (\phi_1 )}{b_{I} (\phi_2 )} \ , \quad
 \Phi = \frac{b_{I\!I} (\phi_3 )}{b_{I\!I} (\phi_2 )} \ ,
\end{eqnarray}
we obtain
\begin{eqnarray}
 S &=& \frac{\ell_I}{2\kappa^2}\int d^4 x \sqrt{-h} \left[ 
   \Psi^2 -1  
   + \alpha  \left( 1- \Phi^2 \right)  \right] R (h) \nonumber \\
   && + \frac{3\ell_I}{\kappa^2} \int d^4 x \sqrt{-h} \left[
    \left(\partial \Psi \right)^2 
   - \alpha \left( \partial \Phi \right)^2 \right] \nonumber\\
   && + \frac{6}{\kappa^2 \ell_I } \int d^4 x \sqrt{-h} \left[ 
   \left( 1 - \beta_1 \right) \Psi^4 
   \nonumber \right.\\
   && \left. - \left( 
   1 - \frac{1}{\alpha }  + \beta_2 \right) 
   -\left(\frac{1}{\alpha} +\beta_3 \right) \Phi^4 \right] \ .
\end{eqnarray}
Here we have defined the dimensionless parameters
\begin{eqnarray}
&&\hspace{-9mm}\alpha = \frac{\ell_{I\!I}}{\ell_I} \ , 
\label{alpha}\\ 
&&\hspace{-1cm}\beta_1 = \frac{\kappa^2 \sigma_1 \ell_I}{6} \ , \quad
\beta_2 = \frac{\kappa^2 \sigma_2 \ell_I}{3} \ , \quad
\beta_3 = \frac{\kappa^2 \sigma_3 \ell_I}{6}\ . 
\label{beta}
\end{eqnarray}
As we have the relation $\ell_{I\!I} < \ell_I$, 
the relation $\alpha <1$ also holds.

After taking the two-brane limit $a(\phi_3 )\rightarrow 0$, 
namely, $\Phi \rightarrow 0$,
we obtain the low energy effective action for the two-brane system. 
\begin{eqnarray}
&& S = \frac{\ell_I}{2\kappa^2}\int d^4 x \sqrt{-h} \left( 
   \Psi^2  + \alpha - 1   \right) R (h) \nonumber \\
   && \quad + \frac{3\ell_I}{\kappa^2} \int d^4 x \sqrt{-h} 
    \left(\partial \Psi \right)^2   \\
   && + \frac{6}{\kappa^2 \ell_I } \int d^4 x \sqrt{-h} \left[ 
   \left( 1 - \beta_1 \right) \Psi^4   - \left( 
   1 - \frac{1}{\alpha }  + \beta_2 \right) \right] \ . \nonumber
\end{eqnarray}
The scalar variable $\Psi$ which describes the distance between two
positive tension branes is called as the radion.


\begin{thebibliography}{99}

\bibitem{Horava}
  P.~Horava and E.~Witten,
  Nucl.\ Phys.\ B {\bf 460}, 506 (1996)
  [arXiv:hep-th/9510209];
  I.~Antoniadis, N.~Arkani-Hamed, S.~Dimopoulos and G.~R.~Dvali,
  Phys.\ Lett.\ B {\bf 436}, 257 (1998)
  [arXiv:hep-ph/9804398];
  A.~Lukas, B.~A.~Ovrut, K.~S.~Stelle and D.~Waldram,
  Phys.\ Rev.\ D {\bf 59}, 086001 (1999)
  [arXiv:hep-th/9803235];
   N.~Arkani-Hamed, S.~Dimopoulos and G.~R.~Dvali,
  Phys.\ Lett.\ B {\bf 429}, 263 (1998)
  [arXiv:hep-ph/9803315].
\bibitem{Dvali}
  G.~R.~Dvali and S.~H.~H.~Tye,
  Phys.\ Lett.\ B {\bf 450}, 72 (1999)
  [arXiv:hep-ph/9812483].
   
\bibitem{Kachru}
S.~Kachru, R.~Kallosh, A.~Linde and S.~P.~Trivedi,
Phys.\ Rev.\ D {\bf 68}, 046005 (2003)
[arXiv:hep-th/0301240];
\\
S.~Kachru, R.~Kallosh, A.~Linde, J.~Maldacena, L.~McAllister and S.~P.~Trivedi,
JCAP {\bf 0310}, 013 (2003)
[arXiv:hep-th/0308055];
 
\bibitem{review}
 R.~Maartens,
  Living Rev.\ Rel.\  {\bf 7}, 7 (2004)
  [arXiv:gr-qc/0312059];
  P.~Brax, C.~van de Bruck and A.~C.~Davis,
  Rept.\ Prog.\ Phys.\  {\bf 67}, 2183 (2004)
  [arXiv:hep-th/0404011];
  S.~Kanno and J.~Soda,
  arXiv:hep-th/0407184.

\bibitem{MWBH}
R.~Maartens, D.~Wands, B.~A.~Bassett and I.~Heard,
Phys.\ Rev.\ D {\bf 62}, 041301 (2000)
[arXiv:hep-ph/9912464].

\bibitem{LMWgw}
D.~Langlois, R.~Maartens and D.~Wands,
Phys.\ Lett.\ B {\bf 489}, 259 (2000)
[arXiv:hep-th/0006007].

\bibitem{KKS}
S.~Kobayashi, K.~Koyama and J.~Soda,
  Phys.\ Lett.\ B {\bf 501}, 157 (2001)
  [arXiv:hep-th/0009160].
\bibitem{HS}  
Y.~Himemoto and M.~Sasaki,
  Phys.\ Rev.\ D {\bf 63}, 044015 (2001)
  [arXiv:gr-qc/0010035].
  
\bibitem{RS1}
L.~Randall and R.~Sundrum,
Phys.\ Rev.\ Lett.\  {\bf 83}, 3370 (1999)
[arXiv:hep-ph/9905221];

\bibitem{KS1}
S.~Kanno and J.~Soda,
Phys.\ Rev.\ D {\bf 66}, 083506 (2002)
[arXiv:hep-th/0207029];

\bibitem{KS2}
S.~Kanno and J.~Soda,
Phys.\ Rev.\ D {\bf 66}, 043526 (2002)
[arXiv:hep-th/0205188];
\\
S.~Kanno and J.~Soda,
Astrophys.\ Space Sci.\  {\bf 283}, 481 (2003)
[arXiv:gr-qc/0209087];
\\
S.~Kanno and J.~Soda,
Gen.\ Rel.\ Grav.\  {\bf 36}, 689 (2004)
[arXiv:hep-th/0303203].

\bibitem{wiseman}
T.~Wiseman,
Class.\ Quant.\ Grav.\  {\bf 19}, 3083 (2002)
[arXiv:hep-th/0201127];
\\
T.~Shiromizu and K.~Koyama,
Phys.\ Rev.\ D {\bf 67}, 084022 (2003)
[arXiv:hep-th/0210066].


\bibitem{LMW}
D.~Langlois, K.~Maeda and D.~Wands,
Phys.\ Rev.\ Lett.\  {\bf 88}, 181301 (2002)
[arXiv:gr-qc/0111013].

\bibitem{TS}
T.~Shiromizu, K.~Koyama and T.~Torii,
  Phys.\ Rev.\ D {\bf 68}, 103513 (2003)
  [arXiv:hep-th/0307151];
K.~Takahashi and T.~Shiromizu,
  Phys.\ Rev.\ D {\bf 70}, 103507 (2004)
  [arXiv:hep-th/0408043];
 P.~Brax and N.~Chatillon,
  JHEP {\bf 0312}, 026 (2003)
  [arXiv:hep-th/0309117];
  P.~Brax and N.~Chatillon,
  Phys.\ Rev.\ D {\bf 70}, 106009 (2004)
  [arXiv:hep-th/0405143].


\bibitem{radion}
P.~Binetruy, C.~Deffayet and D.~Langlois,
Nucl.\ Phys.\ B {\bf 615}, 219 (2001)
[arXiv:hep-th/0101234].

\bibitem{kaloper}
N.~Kaloper,
Phys.\ Rev.\ D {\bf 60}, 123506 (1999)
[arXiv:hep-th/9905210].

\bibitem{BCG}
P.~Bowcock, C.~Charmousis and R.~Gregory,
Class.\ Quant.\ Grav.\  {\bf 17}, 4745 (2000)
[arXiv:hep-th/0007177].


\bibitem{sugumi}
S.~Kanno and J.~Soda,
 Phys. Lett. B 588, 203-209 (2004)
[arXiv:hep-th/0312106];
\\
  P.~L.~McFadden and N.~Turok,
  Phys.\ Rev.\ D {\bf 71}, 021901 (2005)
  [arXiv:hep-th/0409122].
  
\bibitem{Ludo}
L.~Cotta-Ramusino, ``Low energy effective theory for brane world
models'', MPhil thesis, University of Portsmouth (2004).


\bibitem{SMS}
T.~Shiromizu, K.~i.~Maeda and M.~Sasaki,
Phys.\ Rev.\ D {\bf 62}, 024012 (2000)
[arXiv:gr-qc/9910076].
  
\bibitem{nojiri}  
S.~Nojiri, O.~Obregon, S.~D.~Odintsov and V.~I.~Tkach,
  Phys.\ Rev.\ D {\bf 64}, 043505 (2001)
  [arXiv:hep-th/0101003].
  
\bibitem{GenSas}
U.~Gen and M.~Sasaki,
Prog.\ Theor.\ Phys.\  {\bf 105}, 591 (2001)
[arXiv:gr-qc/0011078].

\bibitem{CF}
Z.~Chacko and P.~J.~Fox,
Phys.\ Rev.\ D {\bf 64}, 024015 (2001)
[arXiv:hep-th/0102023].

\bibitem{Contaldi}
C.~R.~Contaldi, L.~Kofman and M.~Peloso,
JCAP {\bf 0408}, 007 (2004)
[arXiv:hep-th/0403270].

\bibitem{Khoury}
J.~Khoury, B.~A.~Ovrut, P.~J.~Steinhardt and N.~Turok,
Phys.\ Rev.\ D {\bf 64}, 123522 (2001)
[arXiv:hep-th/0103239].

\bibitem{Gen}
U.~Gen, A.~Ishibashi and T.~Tanaka,
Phys.\ Rev.\ D {\bf 66}, 023519 (2002)
[arXiv:hep-th/0110286].

\bibitem{Bucher}
M.~Bucher,
arXiv:hep-th/0107148.

\bibitem{KSS}
S.~Kanno, M.~Sasaki and J.~Soda,
Prog.\ Theor.\ Phys.\  {\bf 109}, 357 (2003)
[arXiv:hep-th/0210250].

\bibitem{GT}
  J.~Garriga and T.~Tanaka,
  Phys.\ Rev.\ D {\bf 65}, 103506 (2002)
  [arXiv:hep-th/0112028].
  
\bibitem{Pillado}
  J.~J.~Blanco-Pillado and M.~Bucher,
  Phys.\ Rev.\ D {\bf 65}, 083517 (2002)
  [arXiv:hep-th/0111089].

\bibitem{Neronov}
  A.~Neronov,
  JHEP {\bf 0111}, 007 (2001)
  [arXiv:hep-th/0109090].


\bibitem{Langlois}
  D.~Langlois, R.~Maartens, M.~Sasaki and D.~Wands,
  Phys.\ Rev.\ D {\bf 63}, 084009 (2001)
  [arXiv:hep-th/0012044].
  
\bibitem{WMLL}
D.~Wands, K.~A.~Malik, D.~H.~Lyth and A.~R.~Liddle,
Phys.\ Rev.\ D {\bf 62}, 043527 (2000)
[arXiv:astro-ph/0003278].

\bibitem{Sasaki}
  M.~Sasaki and E.~D.~Stewart,
  Prog.\ Theor.\ Phys.\  {\bf 95}, 71 (1996)
  [arXiv:astro-ph/9507001].

\bibitem{EnqSlo}
K.~Enqvist and M.~S.~Sloth,
Nucl.\ Phys.\ B {\bf 626}, 395 (2002)
[arXiv:hep-ph/0109214].

\bibitem{LW}
D.~H.~Lyth and D.~Wands,
Phys.\ Lett.\ B {\bf 524}, 5 (2002)
[arXiv:hep-ph/0110002].

\bibitem{MT}
T.~Moroi and T.~Takahashi,
Phys.\ Lett.\ B {\bf 522}, 215 (2001)
[Erratum-ibid.\ B {\bf 539}, 303 (2002)]
[arXiv:hep-ph/0110096].

\bibitem{CEW}
E.~J.~Copeland, R.~Easther and D.~Wands,
Phys.\ Rev.\ D {\bf 56}, 874 (1997)
[arXiv:hep-th/9701082].

\bibitem{GS}
J.~Garriga and M.~Sasaki,
Phys.\ Rev.\ D {\bf 62}, 043523 (2000)
[arXiv:hep-th/9912118].

\bibitem{Koyama}
K.~Koyama and J.~Soda,
Phys.\ Lett.\ B {\bf 483}, 432 (2000)
[arXiv:gr-qc/0001033].

\bibitem{KKKK}
K.~Koyama and K.~Koyama,
arXiv:hep-th/0505256.

\bibitem{GPT}
J.~Garriga, O.~Pujolas and T.~Tanaka,
Nucl.\ Phys.\ B {\bf 655}, 127 (2003)
[arXiv:hep-th/0111277];
\\
P.~Brax, C.~van de Bruck, A.~C.~Davis and C.~S.~Rhodes,
Phys.\ Rev.\ D {\bf 67}, 023512 (2003)
[arXiv:hep-th/0209158];
\\
  G.~A.~Palma and A.~C.~Davis,
  Phys.\ Rev.\ D {\bf 70}, 106003 (2004)
  [arXiv:hep-th/0407036];
\\
  J.~L.~Lehners and K.~S.~Stelle,
  Nucl.\ Phys.\ B {\bf 661}, 273 (2003)
  [arXiv:hep-th/0210228].

\bibitem{Soda}
  S.~Kanno and J.~Soda,
  Phys.\ Rev.\ D {\bf 71}, 044031 (2005)
  [arXiv:hep-th/0410061].

\bibitem{Kanno}
  S.~Kanno,
  arXiv:hep-th/0504087.
  
\end{thebibliography}
\end{document}